# An *ab initio* approach to understand the structural, thermophysical, electronic, and optical properties of binary silicide *SrSi₂*: A double Weyl semimetal

Suptajoy Barua[1], B. Rahman Rano[1], Ishtiaque M. Syed[1], S. H. Naqib[2,*]

[1]Department of Physics, University of Dhaka, Dhaka-1000, Bangladesh
[2]Department of Physics, University of Rajshahi, Rajshahi-6205, Bangladesh
*Corresponding author e-mail: salehnaqib@yahoo.com

**ABSTRACT**

A large number of hitherto unexplored elastic, thermophysical, acoustic, and optoelectronic properties of a double Weyl semimetal $SrSi_2$ have been investigated in this study. Density functional theory (DFT) based methodology has been employed. Analyses of computed elastic parameters reveal that $SrSi_2$ is a mechanically stable, ductile, moderately machinable, and relatively soft material. The compound is predicted to be dynamically stable and possesses significant metallic bonding. Study of thermophysical properties, namely, Debye temperature, Grüneisen parameter, acoustic parameters, melting temperature, heat capacity, thermal expansion coefficient, and dominant phonon mode is also indicative of soft nature of $SrSi_2$. The electronic band structure calculations without and with spin-orbit coupling disclose semimetallic character with clear Weyl nodes close to the Fermi level. The electronic dispersion is anisotropic characterized by nearly flat and linear regions within the Brillouin zone. Optical parameters at different photon energies are investigated. $SrSi_2$ shows excellent nonselective reflection spectrum across an extended range of energy encompassing the visible region implying that the compound under study has significant potential to be used as an efficient solar energy reflector. $SrSi_2$ absorbs ultraviolet light quite efficiently. The compound also possesses high refractive index in the low energy. All these optical features can be useful in optoelectronic device applications.

## 1. Introduction

Topological condensed phases of quantum matter have recently been intensively investigated both theoretically and experimentally and have become an increasingly important area of research in condensed matter physics which offers significant potential for device applications [1–4]. Because of the topology of the bulk electronic band structure, topological quantum states exhibit non-trivial surface electronic states [5,6]. Topological materials host topologically protected electronic surface states, which imply that topological transformations keep the basic electronic states of these systems invariant.

Although the starting impetus in studying the topological features of band structures [7,8] arose from the prediction and discovery of two and three-dimensional topological insulators [9–11], intriguing linkages to gapless states have attracted a significant amount of interest in recent years [12,13]. A Weyl semimetal (WSM) is such a gapless topologically protected system whose low energy quasiparticle excitations, known as Weyl fermions, satisfy the Weyl equation. The first prediction of Weyl fermion came in 1929 when Hermann Weyl established the presence of a massless fermion in the Dirac equation [14]. Despite early predictions, the experimental discovery of WSMs happened in recent times [15,16]. Some of the special features of WSMs are the surface Fermi arcs (non-closed Fermi surface), Weyl nodes (always come in pairs with opposite chirality), the chiral anomaly effect etc. Many intriguing properties emerge as a result of these exotic electronic features such as large magnetoresistance, high carrier mobility, light carrier effective mass, novel quantum oscillations and novel superconductivity [17–19]. These fascinating properties of WSMs can be utilized in electronic, spintronic, optical and quantum computing devices.

A non-centrosymmetric binary silicide, $SrSi_2$ with no mirror or inversion symmetry, has recently been predicted to be a quadratic double WSM [20]. A double WSM hosts Weyl nodes possessing chiral charges of ±2. Very recently $SrSi_2$ has been characterized as a type-I WSM via first-principles calculation [21]. The Fermi surface (FS) in a type-I WSM contracts to a point at the Weyl node energy and it respects Lorentz symmetry. In $SrSi_2$, oppositely charged Weyl nodes are positioned at different energies owing to the broken mirror symmetry, which suggests that $SrSi_2$ is an ideal material for investigating the chiral magnetic effect [20]. $SrSi_2$ is a practically favourable compound since it is made up of components that are non-toxic and abundant in nature. Crystallographic and electronic properties of $SrSi_2$ have been studied for a long time [22,23]. $SrSi_2$ crystallizes in a simple cubic structure. Very recently, tunability in the $SrSi_2$ materials class has been studied [24]. It showed that Ca substitution increases the topological metallic feature, whereas Ba doping leads to a fully gapped insulating state. The effect of a uniaxial tensile stress along the z-axis on the double Weyl nodes of $SrSi_2$ was examined recently [25]. Lately, the circular photo-galvanic effect (CPGE) and a connection between Fermi arcs and crystalline handedness in $SrSi_2$ were studied [21]. Various studies on thermoelectric properties of $SrSi_2$ by different groups suggest that it is a suitable choice in thermoelectric conversion technology [26–28]. Some optical and thermodynamic properties for this material were also studied [29]. However, experimental evidence of Weyl nodes in $SrSi_2$ is still waiting.

So far as we know, the mechanical properties of $SrSi_2$ have not been explored to any degree. Moreover, a detailed study of optical properties, as well as thermophysical properties of $SrSi_2$, is still lacking. It should be emphasized that a thorough understanding of the mechanical, thermophysical, and optical properties is essential to probe into the potential applications of the material. With this motivation, our aim in this study is to investigate the bulk properties of $SrSi_2$, such as elastic and optical properties, as well as electronic band structure and electronic energy density of states. In this work, we have studied various elastic moduli, Debye temperature, Grüneisen parameter, several thermophysical quantities, electronic properties related to Fermi surface, along with energy dependent optical parameters of $SrSi_2$. Analyses of the electronic band structure and optical parameters' spectra indicate the semi-metallic nature of SrSi$_2$ in good agreement with one another. In short, our study unmasks several unexplored properties of SrSi$_2$ which will help researchers to comprehend its potential for future applications.

The structure of the rest of the manuscript is as follows: In Section 2, the computational methods are discussed n brief. Section 3 comprises the computational findings of all the properties examined here and their analyses. In Section 4, we finally end with a conclusion and outlook on our study's main findings.

## 2. Computational Methodology

We have carried out all the calculations reported in this work using the CAmbridge Serial Total Energy Package (CASTEP) [30] which implements the Density Functional Theory (DFT) formalism. In the DFT formalism, the many-body system is described based on electron density rather than expressing the system in terms of the complicated electron wave functions and the solution of the Kohn-Sham equation results in the ground state of the crystalline system [31,32].Initially, for the structural properties, both Local Density Approximation (LDA) and Generalized Gradient Approximation (GGA) of Perdew-Burke-Ernzerhof (PBE) scheme were used to model the exchange-correlation effects [33,34]. Because of the localized nature of the trial orbitals, LDA generally underestimates the lattice parameters, whereas GGA overestimates the lattice parameters by relaxing this condition. Norm-conserving pseudopotentials with the Koelling-Harmon relativistic treatment have been utilized to represent the electron-ion interactions [35,36]. The use of norm-conserving pseudopotentials assures that the scattering properties of the pseudopotential have been replicated properly. Optimization of the crystal geometry to obtain the lowest energy structure has been carried out within the Broyden-Fletcher-Goldfarb-Shanno (BFGS) scheme [37]. Valence electronic orbitals: [$4s^2$ $4p^6$ $5s^2$] for Sr and [$3s^2$ $3p^2$] for Si, were used to execute the pseudo atomic computations. An adequate level of total energy convergence has been ensured by adjusting the cut-off energy of the plane-wave basis set to 700 eV. Self-consistent calculations were executed using the density mixing electronic minimizer. A mesh size of 9×9×9 within the Monkhorst-Pack scheme [38] has been employed to sample the special $k$-points of the Brillouin Zone (BZ) of SrSi$_2$. At the same time, mesh sizes of 6×6×6 and 25×25×25 were used to acquire the electronic band structure with spin-orbit coupling (SOC) and the Fermi surfaces, respectively. Several convergence tolerance thresholds, namely, $10^{-5}$ eV/atom for energy, maximum 0.03 eV/Å force, maximum 0.05 GPa stress and highest 0.001 Å displacement, were set to optimize the geometry of the system. A set of prior studies on different classes of materials including topological semimetals demonstrated that SOC has a nominal effect on the bulk structural, elastic, thermophysical and optical properties calculations [39–42]. Accordingly, SOC has not been considered in these studies except in the band structure calculations of SrSi$_2$ that includes SOC to reveal the topological aspects with clarity.

The *stress-strain* method from the CASTEP code has been used to compute the single-crystal elastic constants, $C_{ij}$ and elastic compliances, $S_{ij}$. All the other elastic and thermophysical properties mentioned in this study have been derived with the help of these constants. Fermi surface, electronic band structure and electronic energy density of states calculations were done for the geometrically optimized structure of SrSi$_2$.

While conducting the optical properties calculations, the complex dielectric function, $\varepsilon(\omega) = \varepsilon_1(\omega) + i\varepsilon_2(\omega)$, was evaluated first. Considering the matrix elements of the electronic transitions between occupied and unoccupied electronic states, the imaginary part, $\varepsilon_2(\omega)$, was determined using the formula supported by CASTEP, which is defined as:

$$\varepsilon_2(\omega) = \frac{2e^2\pi}{\Omega\varepsilon_0} \sum_{k,v,c} |\Psi_k^c|\hat{u}.\vec{r}|\Psi_k^v|^2 \delta(E_k^c - E_k^v - E) \tag{1}$$

In this formula, $\Omega$ represents the unit cell volume, $\omega$ denotes the incident photon's angular frequency, $e$ refers to an electron's charge, the polarization of the incident electric field is defined by the unit vector $\hat{u}$, $\Psi_k^c$ and $\Psi_k^v$ are the electron wave functions corresponding to the conduction and valence band that have crystal momentum $\hbar k$, respectively. The delta function assures the energy and momentum conservation in the optical transition process. The real part of the dielectric function, $\varepsilon_1$ was obtained by applying the Kramers-Kronig transformation to the corresponding imaginary part, $\varepsilon_2$ bearing in mind that a causal response is characterized by the dielectric function. Complete information of the complex dielectric function was utilized to compute the other optical properties shown in this study using well established methodology [43].

## 3. Results and Analysis

### *3.1. Structural features of SrSi$_2$*

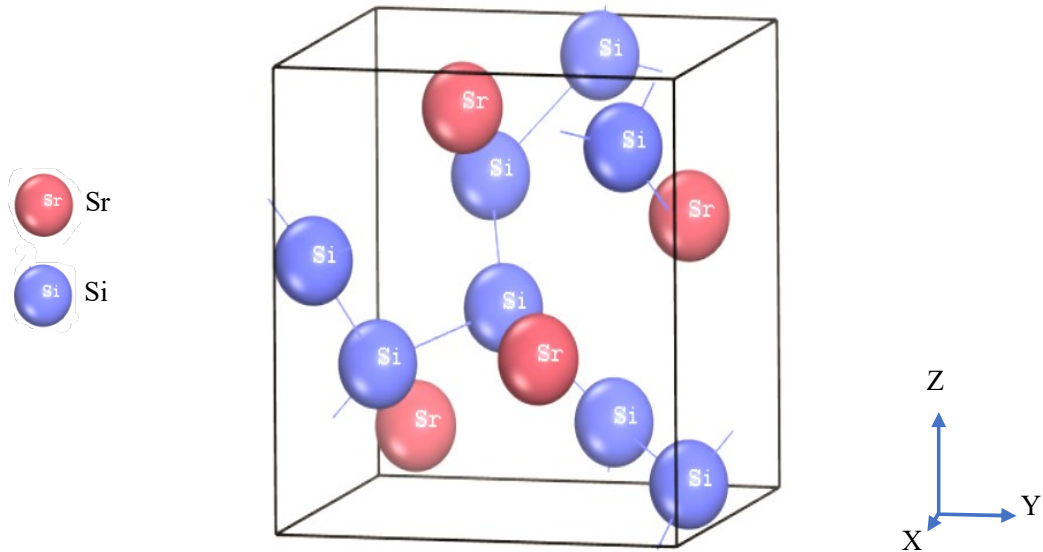

**FIG. 1.** Schematic crystal structure of *SrSi$_2$*.

The lattice system of SrSi$_2$ has a cubic structure associated with the space group $P4_332$ (No. 212) [22,23]. The unit cell crystal structure of the SrSi$_2$ system is schematically illustrated in Fig. 1. The lattice parameter and atomic position optimization process in our computation was initiated with the experimental lattice parameters and atomic positions [23]. As it is seen from figure 1, the unit cell of the SrSi$_2$ crystal structure contains 4 Sr atoms and 8 Si atoms occupying the Wyckoff positions 4a (1/8, 1/8, 1/8) and 8c (0.4223, 0.4223, 0.4223), respectively [23,25]. Therefore, the unit cell of SrSi$_2$ comprises four formula units.

Along with the previously reported experimental values, the optimized lattice parameters and cell volumes of SrSi$_2$, obtained from both GGA and LDA calculations are listed in Table I. As anticipated, GGA and LDA calculations respectively overestimate and underestimate the structural parameters. From Table I, it is clear that the crystal parameters obtained from the GGA calculations are more consistent with the experimental values than those obtained from the LDA calculations. For the reliability of *ab-initio* calculations, using proper lattice parameters is a vital factor. Therefore, the results of the GGA calculations have been presented in the following sections since it provides better valuations of the ground state structural parameters of SrSi$_2$.

**TABLE I:** The optimized lattice parameters and cell volumes of SrSi$_2$ obtained from GGA and LDA calculations. The lattice parameters *a*, *b*, and *c* are in Å and the unit cell volume (*V*) is in Å$^3$.

| Functional (Reference) | a | b | c | V |
|---|---|---|---|---|
| *GGA* (This work) | 6.553 | 6.553 | 6.553 | 281.38 |
| *LDA* (This work) | 6.412 | 6.412 | 6.412 | 263.66 |
| *Experimental* [23] | 6.535 | 6.535 | 6.535 | 279.09 |

### 3.2. Elastic properties of SrSi$_2$

The investigation of a material's elastic properties is an important research field for the advancement of technology since it provides the necessary information to understand the behaviour of the material under mechanical stress of various types. Particularly, the elastic constants of a crystal arise in relation to the mechanical and dynamical properties calculations. The elastic constants of a crystal are usually put in a 6 × 6 symmetric matrix known as the stiffness matrix or the elastic matrix. The number of independent components in that matrix depends on the number of symmetry constraints present in the crystal class. Since cubic systems exhibit a high degree of symmetry, the stiffness matrix of a cubic crystal is composed of just three distinct elastic constants, namely, $C_{11}$, $C_{12}$ and $C_{44}$. Therefore, SrSi$_2$ has three single-crystal elastic constants that are independent of each other, i.e., $C_{11}$, $C_{12}$ and $C_{44}$, as it crystallizes into a cubic structure. The computed single-crystal elastic constants of SrSi$_2$, as well as the obtained elastic compliance constants, have been tabulated in Table II.

Necessary and sufficient conditions to ensure the mechanical stability of crystalline structures under static stress are concisely expressed in the form of inequalities known as the *Born stability criteria* [44,45]. Born stability criteria take different forms for different classes of crystals [45]. In the case of cubic crystals, the necessary and sufficient conditions ensuring the mechanical stability of the structure, i.e., the Born stability criteria, reduce to the following form [45]:

$$C_{11} - C_{12} > 0, \qquad C_{11} + 2C_{12} > 0, \qquad C_{44} > 0 \qquad (2)$$

The computed elastic constants of SrSi$_2$ fulfill all the inequalities in Born stability criteria suggesting that the system under consideration is mechanically stable.

The first three diagonal elastic constants, $C_{11}$, $C_{22}$ and $C_{33}$, reflect the capability of a material to withstand tensile stress applied in the crystallographic *a-*, *b-* and *c-*directions, respectively. $C_{11} = C_{22} = C_{33}$ for the cubic crystal SrSi$_2$ indicates that the bonding strengths along the crystallographic directions are identical. The last three diagonal elastic constants, $C_{44}$, $C_{55}$ and $C_{66}$, are used to determine a material's response to shearing stress. For instance, $C_{44}$ measures the resistance against shear due to a tangentially applied stress across the (100) plane in the [010] direction. $C_{44}$ also represents the indentation hardness of a material. The low value of $C_{44}$ ($C_{44} = C_{55} = C_{66}$) compared to $C_{11}$ in SrSi$_2$ implies that the crystal is less resistant to shearing strain. Therefore, it is anticipated that the mechanical failure mode of SrSi$_2$ would be controlled by the shearing strain, rather than the unidirectional strains. The shear components that are off-diagonal, i.e., $C_{12}$, $C_{13}$ and $C_{23}$, represent the resistance of a crystal against volume conserving orthogonal distortions. In the case of SrSi$_2$, the value of $C_{12}$ ($C_{12} = C_{13} = C_{23}$) is lower than the value of $C_{11}$ which signifies the weaker resistance of SrSi$_2$ against orthogonal distortions.

The tetragonal shear modulus, $C' \left(= \frac{C_{11} - C_{12}}{2}\right)$ is a useful parameter to determine the dynamical stability of a crystalline solid [46,47]. It also gives an idea about the stiffness of a crystal. Furthermore, $C'$ is connected with the slow transverse acoustic waves and plays a vital role in the structural transformations of a system [48]. The dynamical stability of a crystal is ensured if the value of $C'$ is positive whereas the negative value of $C'$ indicates that the material is dynamically unstable. The calculated value of $C'$ for SrSi$_2$ is listed in Table II. A positive $C'$ implies that SrSi$_2$ is dynamically stable.

We have also calculated the Kleinman parameter, $\zeta$, that describes a compound's stability under bending and stretching. The Kleinman parameter is dimensionless and usually attains a value between 0 and 1. The Kleinman parameter qualitatively represents the contribution of bond bending and bond stretching to the mechanical strength of the material. Moreover, the relative shift of the cation and anion sub-lattice positions owing to volume conserving distortions in which the atomic positions are not set by the crystal symmetry can be explained using the Kleinman parameter [49]. The value of $\zeta$ for SrSi$_2$ has been evaluated using the following relation [50]:

$$\zeta = \frac{C_{11} + 8C_{12}}{7C_{11} + 2C_{12}} \tag{3}$$

The calculated value of the Kleinman parameter for SrSi$_2$ is listed in Table II. The value of $\zeta$ for SrSi$_2$ is 0.665 which implies that the bond bending contribution is dominating over the bond stretching contribution in the mechanical strength of SrSi$_2$.

We have further calculated the Cauchy pressure, $C''(= C_{12} - C_{44})$ that reflects the nature of bonding in a compound at the atomic level [51,52]. A material with positive Cauchy pressure is typically characterized to be a ductile material while the material's brittleness is represented by a negative value [53]. As stated by *Pettifor* [54], strong metallic non-directional bonding is observed in materials with a large positive value of $C''$ while conversely, an angular character in the bonding corresponds to the negative value of $C''$. The value of the Cauchy pressure becomes zero for a bonding that can be explained by simple pairwise potentials, like the Lennard-Jones potential. Cauchy pressure as calculated for SrSi$_2$ is presented in Table II. The positive value of $C''$ for SrSi$_2$ suggests the ductile nature of SrSi$_2$ as well as the existence of metallic bonding in the compound [54].

**TABLE II:** The single crystal elastic constants ($C_{ij}$ in GPa) and elastic compliance constants ($S_{ij}$ in 1/GPa), tetragonal shear modulus ($C'$ in GPa), Kleinman parameter ($\zeta$) and Cauchy pressure ($C''$ in GPa) for *SrSi$_2$*.

| $C_{11}$ | $C_{12}$ | $C_{44}$ | $S_{11}$ | $S_{12}$ | $S_{44}$ | $C'$ | $\zeta$ | $C''$ |
|---|---|---|---|---|---|---|---|---|
| 83.17 | 45.60 | 42.14 | 0.020 | -0.007 | 0.024 | 18.79 | 0.665 | 3.46 |

To further elucidate the ductility of SrSi$_2$, we have calculated the polycrystalline elastic moduli and summarized those in Table III. The elastic moduli of polycrystalline aggregates can be computed from the single-crystal elastic constants by utilizing various approximations [55-57]. In case of the Voigt approximation, an imbalance in the actual stresses among grains due to the discontinuous stress distribution inside the grains gives rise to the upper bound of the polycrystalline elastic moduli [58]. The Reuss approximation, on the other hand, considers a continuous stress distribution while the strain is distributed discontinuously inside the grains. This discontinuity leads to the lower bound of the polycrystalline elastic moduli [59]. However, the prescription that provides the closest approximation to the true polycrystalline elastic moduli, known as Hill's approximation, calculates the arithmetic mean of the two bounds obtained from the respective Voigt and Reuss approximations [60]. The bulk modulus, $B$ and the shear modulus, $G$ of a material represent the capability of the material to withstand volume changes caused by isotropically applied pressure and plastic deformation due to shear, respectively. As anticipated from the elastic constant calculations, the lower value of the shear modulus relative to the bulk modulus indicates that the applied shear component can control the mechanical failure of SrSi$_2$. Low values of the elastic moduli of SrSi$_2$ compared to several other binary metallic compounds [41,42,61,62], indicate its soft nature. The bulk to shear modulus ratio of a solid, known as Pugh's ratio, $B/G$ is used to understand the ductile or brittle nature of the material [40,42,57,63]. The critical value of $B/G$ at the interface between ductile and brittle material is about 1.75. Ductility is associated with a high value of the Pugh's ratio (> 1.75) while solids with a low value (< 1.75) exhibit brittleness [57,63]. Therefore, the calculated Pugh's ratio for SrSi$_2$ reveals the compound's ductile character, which is in accordance with the Cauchy pressure. Young's modulus, $E$ is defined as the ratio of the tensile stress to the longitudinal strain that measures the stiffness of a material [55]. For SrSi$_2$, the low value of $E$ suggests that it cannot endure high tensile stress. The value of $E$ has been evaluated from the bulk and shear modulus by applying the following formula [55,57]:

$$E = \frac{9BG}{3B + G} \quad (4)$$

The Poisson's ratio, $v$ is a widely used parameter to describe several features such as ductility/brittleness, compressibility and characteristics of bonding force of a material. The Poisson's ratio also represents a compound's stability against shear. Better stability against shear is predicted if the value of $v$ is comparatively smaller. The numerical value of $v$ typically lies in the range, $-1.0 \leq v \leq 0.50$. The critical value for $v$ at the boundary between ductile and brittle polycrystalline material is about 0.26 [64,65]. A material is suggested to be ductile (brittle) if the value of $v$ is greater (less) than 0.26. Therefore, the value of $v$ for SrSi$_2$, as displayed in Table III, indicates the ductile nature of SrSi$_2$. This prediction is in complete agreement with the previously analyzed elastic parameters in this study. Commonly, the Poisson's ratio resides within 0.25 and 0.50 for solids where the interatomic interactions are dominated by central forces [66]. Moreover, the value of $v$ in pure covalent compounds is around 0.10 while it is around 0.33 for metallic compounds. This implies that metallic bonding is prominent in SrSi$_2$. The value of $v$ has been evaluated using the following expression [55,57]:

$$v = \frac{3B - 2G}{2(3B + G)} \quad (5)$$

The study of machinability index, $\mu_M (= B/C_{44})$ has attracted significant research interest in the manufacturing industry due to its capability of predicting the convenience of applying various cutting tools to machine material [68,69]. The machinability of a material is determined by the properties of the working material and the cutting method used to machine the material. A high value of the machinability index usually implies better dry lubricating properties, low machining cost and low feed force. The calculated value of $\mu_M$ is listed in Table III. Along with $\mu_M$, we have also calculated the hardness, $H$ of SrSi$_2$ since it provides a reliable guide to machinability. Materials with a high value of $H$ are important for cutting tools and wear-resisting coatings albeit they are comparatively difficult to machine. Hardness of SrSi$_2$ has been calculated using Eqn. 6 and is displayed in Table III [39,41].

$$H = \frac{(1 - 2v)E}{6(1 + v)} \quad (6)$$

**TABLE III:** The isotropic bulk modulus (*B* in GPa) and shear modulus (*G* in GPa) for polycrystalline *SrSi₂* obtained from the single crystal elastic constants using Voigt, Reuss and Hill's approximations. The Pugh's ratio (*B/G*), Young's modulus (*E* in GPa), Poisson's ratio (*v*), machinability index ($\mu_M$) and hardness (*H* in GPa) for *SrSi₂* estimated from Hill's approximation.

| $B_R$ | $B_V$ | $B_H$ | $G_R$ | $G_V$ | $G_H$ | $B/G$ | $E$ | $v$ | $\mu_M$ | $H$ |
|---|---|---|---|---|---|---|---|---|---|---|
| 58.13 | 58.13 | 58.13 | 28.15 | 32.80 | 30.47 | 1.907 | 77.82 | 0.277 | 1.379 | 4.535 |

The results displayed in Table III imply that SrSi₂ is a compound with moderate hardness and machinability. The polycrystalline elastic moduli are also quite low compared to many other metallic binary compounds [70-72].

### 3.3. Thermophysical properties of SrSi₂

#### 3.3.1. Acoustic behavior

A material's elastic characteristics have a strong influence on how elastic waves propagate through it. Moreover, electrical and thermal conductivity largely depend on the acoustic behaviour of the material. In the quantum theory of solids, the relation, $k \propto Clv$, where heat capacity of a dielectric which is the same as that of a phonon gas, is denoted by *C*, *v* stands for the average phonon velocity, which is about the same as the sound velocity, and mean free path of phonons is represented by *l*, directly links the sound velocity in a crystal to its thermal conductivity, *k*. For metallic systems, conduction electrons also contribute to the thermal conductivity. Investigating the nature of elastic waves has become a crucial research field in geology, materials science, musical instrument design, seismology and medical sciences. The transverse, $v_t$ and longitudinal, $v_l$, elastic wave velocities travelling through a crystalline solid can be formulated based on the polycrystalline elastic moduli as follows [67]:

$$v_t = \sqrt{\frac{G}{\rho}} \tag{7}$$

and,

$$v_l = \sqrt{\frac{B + \frac{4G}{3}}{\rho}} \tag{8}$$

The density of the material is referred by the symbol $\rho$.

The average sound velocity, $v_m$ in the crystal can be further computed from these transverse and longitudinal wave velocities from the following relation [68]:

$$v_m = \left[\frac{1}{3}\left(\frac{2}{v_t^3} + \frac{1}{v_l^3}\right)\right]^{-\frac{1}{3}} \quad (9)$$

In Table IV, the corresponding acoustic wave velocities computed for SrSi$_2$ are presented.

Proper analysis of the acoustic impedance of a material is required in transducer design, aircraft engine design, automobiles and many underwater acoustic applications. A material's acoustic impedance is a measure of the acoustic energy that transmits between two media. The difference in acoustic impedances of two media gives a measurement of how much energy is reflected and transmitted upon arrival of a sound wave at their interface. Sound wave is mostly transmitted if the difference is nearly zero whereas most of it gets reflected when the difference is large. Acoustic impedance characterizes how the applied sound pressure is related to the resulting particle velocity. When a constant sound pressure is applied, particle velocity is inversely related to the acoustic impedance of a medium. The following equation defines the acoustic impedance of a material in terms of the shear modulus and the density of the material [73]:

$$Z = \sqrt{\rho G} \quad (10)$$

The above equation implies that the acoustic impedance should vary with temperature since both the shear modulus and density are temperature dependent. Table IV includes the value of $Z$ as computed for SrSi$_2$.

Another important design parameter utilized in the design of soundboards is the intensity of the sound radiation. The intensity, I, can be expressed as a proportionality relation that scales with the shear modulus and the density of a material in the following manner [73]:

$$I \propto \sqrt{\frac{G}{\rho^3}} \quad (11)$$

The term, $\sqrt{\frac{G}{\rho^3}}$ in the above relation, known as the *radiation factor*, is used in the selection of suitable materials for soundboards design. The effectiveness of a vibrating surface to function as a sound radiator can be determined by the radiation factor. Typically, a high value of it is preferable for soundboards. For example, since the radiation factor of spruce is high (8.6 m$^4$/kg-s), it is the most commonly used material in the front plate of violins, while maple, with about half the radiation factor of spruce (5.4 m$^4$/kg-s), is basically used in the back plate, which serves to reflect. Furthermore, the radiation factor describes how the propagation of structural waves in vibrating elements couples with the noise generated by the structure. The radiation factor calculated for SrSi$_2$ is shown in Table IV. The computed values of Z and $\sqrt{\frac{G}{\rho^3}}$ suggest that the compound under study should be a very efficient reflector of acoustic energy.

**TABLE IV:** The density ($\rho$ in g/cm³), longitudinal ($v_l$), transverse ($v_t$) and average ($v_m$) elastic wave velocity (in m/s), the acoustic impedance (Z in Rayl) and the radiation factor ($\sqrt{G/\rho^3}$ in m⁴/kg-s) of *SrSi₂* obtained from the polycrystalline elastic modulus.

| $\rho$ | $v_l$ | $v_t$ | $v_m$ | $Z (\times 10^6)$ | $\sqrt{G/\rho^3}$ |
|---|---|---|---|---|---|
| 3.394 | 5394.2 | 2996.5 | 3337.4 | 10.17 | 0.88 |

*3.3.2. Debye temperature and Grüneisen parameter*

Many important thermophysical properties such as specific heat, thermal conductivity, phonon dynamics, charge transport and melting temperature of a material are closely related to an important lattice dynamical and thermal parameter, called Debye temperature. According to the Debye model, Debye temperature, $\theta_D$, signifies the highest achievable energy due to a single normal vibration of a crystal. The formula $\theta_D = \hbar\omega_D/k_B$ relates the Debye temperature to the highest allowed phonon frequency, $\omega_D$. Debye temperature contains information about the bonding strength among the atoms in a crystal. When the value of $\theta_D$ is low, it suggests the softness of the material. Furthermore, $\theta_D$ is related to the superconducting transition temperature because, for the phonons associated with conventional superconductors, the cut-off value of boson energy in Cooper pairing is set by Debye temperature. A boundary between the classical and quantum behaviour of solids is also marked by the Debye temperature. When the system temperature is low, determining $\theta_D$ either from the elastic constants or specific heat measurements is equivalent since the vibrational excitations in the low-temperature region emerge purely from the acoustic modes. Accordingly, in our study, we have used one of the standard methods for determining $\theta_D$, based on the elastic constants as follows [68]:

$$\theta_D = \frac{h}{k_B}\left[\left(\frac{3n}{4\pi}\right)\frac{N_A\rho}{M}\right]^{\frac{1}{3}} v_m \quad (12)$$

where *n* denotes the number of atoms present in the molecule, $\rho$ represents the density, $N_A$ stands for the Avogadro's number, *M* denotes the molecular weight, $k_B$ defines the Boltzmann's constant, *h* defines the Planck's constant, and the average sound velocity is expressed as $v_m$. The value of the Debye temperature for our material (listed in Table V) is low which is suggesting the softness of SrSi₂ as anticipated from the elastic moduli calculations.

Also, the Grüneisen parameter, $\gamma$, for SrSi₂ has been calculated and presented in Table V. The Grüneisen parameter is a valuable dimensionless quantity that measures the lattice anharmonicity. High level of anharmonicity is observed for large values of $\gamma$. It contains information on how phonon frequencies and phonon damping depend on temperature. It also represents how phonon frequency is related to volume change resulting from the anharmonicity in the lattice potential. Moreover, this parameter is helpful in understanding the thermal expansion effects in a crystal. The Grüneisen parameter is widely used for investigating the phase transitions related to volume changes. We utilized the following formula which is based on the Poisson's ratio to compute the Grüneisen parameter [74]:

$$\gamma = \frac{3(1+\nu)}{2(2-3\nu)} \quad (13)$$

The computed Grüneisen parameter of SrSi$_2$ is 1.64. This value is typical to many other compounds having intermediate level of anharmonicity [42,72].

### 3.3.3. Melting temperature

Determination of melting temperature, $T_m$ is a vital component in the study of a material since it gives an idea of the temperature domain over which the material can be used in applications. Melting temperature of a crystalline solid is deeply connected to its bonding energy and thermal expansion. Generally, high bonding energy, high cohesive energy, and low thermal expansion are observed in materials having high melting temperatures [75]. In this work, the elastic constants were utilized to estimate the melting temperature by using the relation expressed as [75]:

$$T_m = 354K + (4.5K/GPa)\left(\frac{2C_{11}+C_{33}}{3}\right) \pm 300K \quad (14)$$

Melting temperature of SrSi$_2$ is listed in Table V. Compared to many other metallic compounds [70-72,76] the melting temperature of SrSi$_2$ is low and the compound under study is not suitable for high temperature applications.

### 3.3.4. Thermal expansion coefficient

The thermal expansion coefficient, $\alpha$, gives a measure of the thermal strain per degree of temperature change. The thermal expansion coefficient is linked with various physical properties of a material, such as the specific heat, thermal conductivity and electron/hole effective mass. The study of thermal expansion coefficient of a material is useful in estimating the potentiality of the system as thermal barrier coating. The thermal expansion coefficient of a material has a significant influence on the epitaxial growth of crystals and the reduction of damaging effects when used in spintronic and electronic devices. Low value of $\alpha$ of a material indicates its stability over a broad range of temperatures. The thermal expansion coefficient has been calculated by means of the following equation [73]:

$$\alpha = \frac{1.6 \times 10^{-3}}{G} \quad (15)$$

In the above equation, $G$ denotes the isothermal shear modulus (in GPa). Empirically, $\alpha$ is related to the melting temperature via the relation: $\alpha \approx 0.02/T_m$ [73]. The value of $\alpha$ for SrSi$_2$ is tabulated in Table V. In SrSi$_2$, high value of $\alpha$ implies that volume changes rapidly with temperature. Therefore, the compound under study is suitable for applications as sensors where a high thermal expansion coefficient is needed. The value of $\alpha$ for SrSi$_2$ obtained from the empirical relation: $\alpha \approx 0.02/T_m$ is $2.75 \times 10^{-5} K^{-1}$ which is slightly lower than that obtained from Eqn. 15.

### 3.3.5. Heat capacity and dominant phonon wavelength

The amount of heat required for a particular amount of change in an object's temperature is set by a physical quantity known as the heat capacity, $C_P$. Materials with large heat capacity usually exhibit high thermal conductivity and low thermal diffusivity. Although heat capacity is an extensive property of a substance, the corresponding heat capacity per unit volume, often called the volumetric heat capacity, is an intensive property. Volumetric heat capacity is often expressed as the multiplication of the specific heat capacity and density of the material, i.e., $\rho C_P$. This parameter measures the thermal energy that must be provided in order to raise the temperature of an object with unit volume by one degree. $\rho C_P$ can be evaluated via the equation given by [73]:

$$\rho C_P = \frac{3k_B}{\Omega} \qquad (16)$$

where $N = 1/\Omega$ denotes the atomic number density. A higher value of $\rho C_P$ in heat transmission indicates that the system will take longer to reach equilibrium. The calculated value of $\rho C_P$ for SrSi$_2$ is also presented in Table V. Low value of $\rho C_P$ for SrSi$_2$ compared to many other metallic substances [41,61,62] suggests that the compound under study is comparatively more sensitive to changes in thermal energy.

We also evaluated the dominant phonon wavelength of SrSi$_2$ since phonons, i.e., the quantum of lattice vibration, play a vital role in thermal and transport properties of a material. By definition, the peak of the phonon distribution curve is located at the dominant phonon wavelength, $\lambda_{dom}$. Large values of the dominant phonon wavelength are typically observed in compounds having low density, high shear modulus and high mean sound velocity. The existence of coherence effects can be assessed by comparing $\lambda_{dom}$ to the system dimensions [77]. The value of $\lambda_{dom}$ varies inversely with temperature and can be evaluated from the relationship given by [78]:

$$\lambda_{dom} = \frac{12.566 v_m}{T} \times 10^{-12} \qquad (17)$$

where, the mean speed of sound is denoted by $v_m$, which is in ms$^{-1}$, and the temperature is denoted by $T$, which is in degree Kelvin. At 300 K, the value of $\lambda_{dom}$ is usually in the order of 1 nm, and at around 1 K, it is around 100 nm for crystalline solids [77]. The value of $\lambda_{dom}$ calculated for SrSi$_2$ at 300 K is shown in Table V.

**TABLE V:** The Debye temperature ($\theta_D$ in K), Grüneisen parameter ($\gamma$), melting temperature ($T_m$ in K), thermal expansion coefficient ($\alpha$ in K$^{-1}$), heat capacity per unit volume ($\rho C_P$ in J/m$^3$-K) and the dominant phonon wavelength at 300K ($\lambda_{dom}$ in m) of *SrSi$_2$*.

| $\theta_D$ | $\gamma$ | $T_m(\pm 300)$ | $\alpha\ (\times 10^{-5})$ | $\rho C_P (\times 10^6)$ | $\lambda_{dom}(\times 10^{-12})$ |
|---|---|---|---|---|---|
| 347.29 | 1.64 | 728.28 | 5.25 | 1.77 | 139.8 |

## 3.4. Electronic properties of SrSi$_2$

### 3.4.1. Electronic band structure

The bulk electronic band structure contains plenty of information that is useful to explain various physical properties like optical and transport properties of the material. We have revisited the electronic band structure of SrSi$_2$ without spin-orbit coupling and with its inclusion in this section. The electronic band structure of SrSi$_2$ along a particular path joining the points of high symmetry within the Brillouin zone (BZ) is illustrated in figure 2. Figure 2(a) depicts the band structure in the absence of SOC while the band structure plot in figure 2(b) includes SOC. The Fermi level, $E_F$ is set at zero eV in these plots. A crossing between the valence and conduction bands along the Γ-X direction is observed in the band structure of SrSi$_2$. The extent of this band crossing characterizes SrSi$_2$ as a semimetal. In the absence of SOC, the bands disperse almost linearly in the vicinity of the crossings along the Γ-X direction. The crossings occur at two distinct points in the Γ-X line. Previously reported results reveal that without SOC these two points are Weyl nodes with chiral charges of ±1 whereas with SOC these two points are Weyl nodes with chiral charges of ±2 [20,21,24]. This suggests that SrSi$_2$ is a double-node WSM. Upon inclusion of SOC, the bands with different rotational eigenvalues stay gapless, whereas bands become gapped out if they possess the same rotational eigenvalues. The electronic band structure of SrSi$_2$ reveals that oppositely charged Weyl nodes are positioned at different energies. Probably, this property emerges owing to the broken mirror symmetry in SrSi$_2$. Such an interesting property facilitates SrSi$_2$ to be a perfect material for realizing the chiral magnetic effect. These results agree very well with the prior studies [20,21,24]. Therefore, the overall band structure features of SrSi$_2$ uncover a lot of information about the material and suggest the topological signature of SrSi$_2$. The bands along X-M close to the Fermi level are almost flat. The charge carriers in these bands are expected to possess significantly higher effective mass.

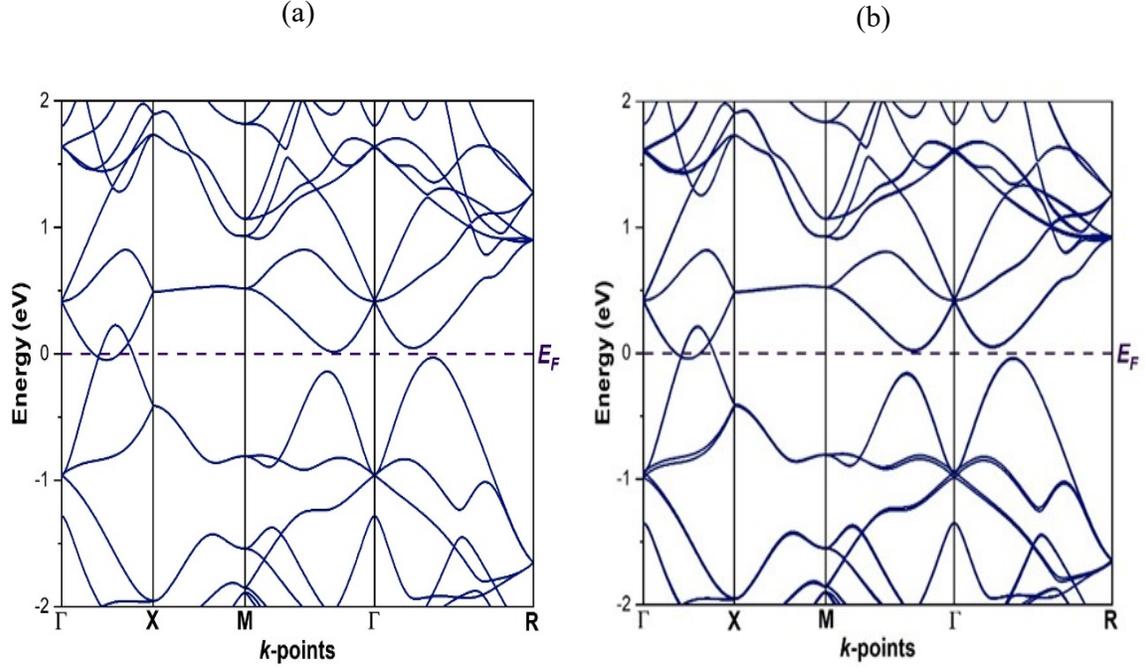

**FIG 2.** The electronic band structure (a) without SOC and (b) with SOC, of *SrSi₂* along the high symmetry directions in the BZ.

*3.4.2. Density of states*

Figure 3 depicts the total density of states (TDOS) of SrSi$_2$calculatedfor the geometrically optimized structure as a function of energy, (E-E$_F$), measured using the Fermi level as a reference. The Fermi level, E$_F$ is shown by a vertical solid line at 0 eV. The metallic character of SrSi$_2$ is indicated by the non-zero TDOS value at the Fermi level, which coherently accords with our electronic band structure estimates. A pseudogap located almost at the Fermi level is found in the TDOS plot of SrSi$_2$. Commonly, a gap or deep valley between the bonding and anti-bonding peaks in traditional metallic systems is known as the pseudogap that is connected to a compound's electronic stability [57,79,80]. The existence of a pseudogap almost at the Fermi level suggests high structural stability of SrSi$_2$ [81].The bonding and anti-bonding peaks are within 2 eV from E$_F$. Accordingly, chemical or mechanical processes, for instance, doping or pressure, can be employed to adjust the Fermi level across these peaks and therefore, major modifications in the electronic property of SrSi$_2$can be implemented. As per calculations for SrSi$_2$, the TDOS value at the Fermi level is 2.45 states/eV-unit cell. This value of TDOS at Fermi level, N(E$_F$) is associated with the electronic stability of a compound [82,83]. Besides, N(E$_F$) can be utilized to determine the electron-electron interaction parameter, often called the repulsive Coulomb pseudopotential, $\mu^*$, by means of the following equation [84]:

$$\mu^* = \frac{0.26 N(E_F)}{1 + N(E_F)} \tag{18}$$

Since N($E_F$) for SrSi$_2$ is 2.45 states/eV-unit cell, the repulsive Coulomb pseudopotential, $\mu^*$ of SrSi$_2$ is found to be 0.18. The formation of Cooper pairs necessary for superconductivity is hindered by this repulsive Coulomb pseudopotential which results in the lowering of the transition temperature, T$_C$ of superconducting materials [85-87].

Alongside TDOS, the atomic orbital resolved partial density of states (PDOS) of SrSi$_2$ is also illustrated in figure 3 to further elucidate the contributions of each atom in the TDOS. It is evident that near E$_F$, the majority of the contributions in the TDOS of SrSi$_2$ come from the Sr-4p and Si-3p electronic orbitals.

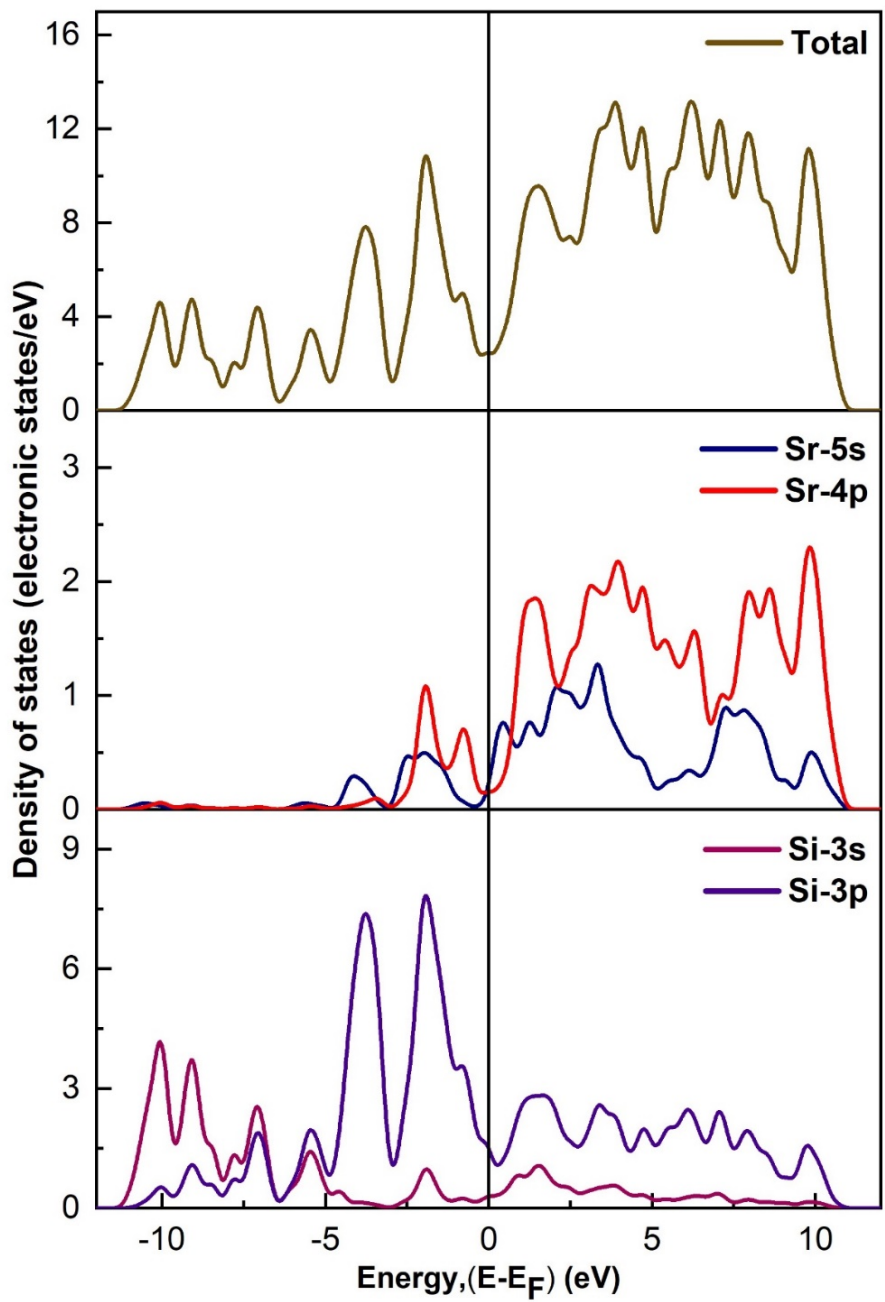

**FIG 3.** Total and partial density of states for *SrSi₂*.

*3.4.3. Fermi surface*

A proper investigation of Fermi surface (FS) is a useful tool for realizing the thermal, electrical, magnetic and optical properties of metals and semi-metals [88]. The isoenergy surface that divides the occupied and unoccupied states in the reciprocal space at zero temperature is defined as the Fermi surface. FS gives information about the behaviour of occupied and unoccupied electronic states in a metallic system in the low-temperature region. Several quantum phenomena including superconductivity, topological insulation and ferromagnetism are largely influenced by the topology of the Fermi surface. Particularly, it can be applied to anticipate the complex behaviour of systems without caring too much about the computational details. The calculated FS plot of SrSi$_2$ is shown in figure 4. Among the 44 bands found in the electronic structure calculations of SrSi$_2$, only two bands cross the Fermi level which coherently agrees with the band structure calculations in this study. The Fermi surfaces of SrSi$_2$ are constructed from these two bands. From the FS plot, it is evident that SrSi$_2$ contains both hole-like and electron-like sheets. Hole-like sheets along the Γ-X line are observed for one of the bands. For the other band, an electron-like sheet appears surrounding the Γ point.

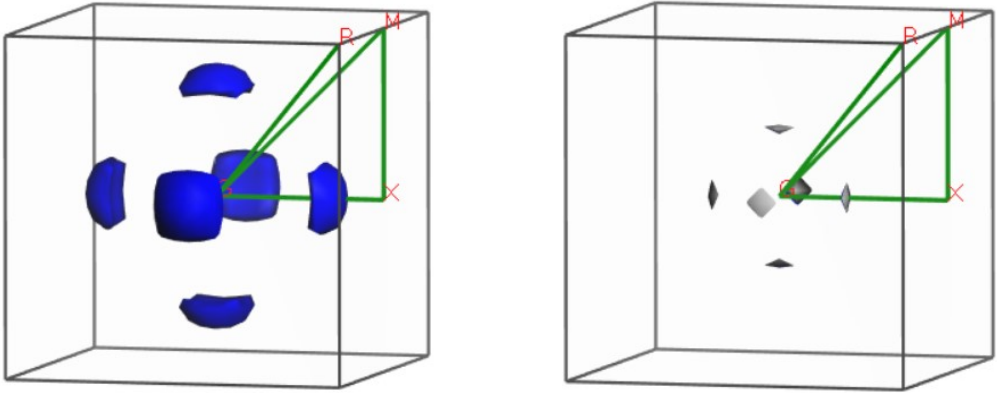

**FIG 4.** Fermi surface of *SrSi$_2$* for the two bands crossing the Fermi level.

*3.5. Optical properties*

The response of a material when irradiated with electromagnetic radiations depends on the optical properties. Therefore, a detailed investigation into the optical properties of a material is essential to explore its utility in optoelectronic applications. Furthermore, the behaviour of optical spectra is closely related to the electronic band structure and the electronic density of states of the system. In this work, we have evaluated several energy-dependent optical parameters for SrSi$_2$, namely the complex dielectric function $(real, \varepsilon_1(\omega)\ and\ imaginary, \varepsilon_2(\omega))$; the real part, $n(\omega)$ and the imaginary part (often called the extinction coefficient), $k(\omega)$ of the refractive index; the optical conductivity $(real, \sigma_1(\omega)\ and\ imaginary, \sigma_2(\omega))$; the reflectivity, $R(\omega)$; the absorption coefficient, $\alpha(\omega)$, and the loss function, $L(\omega)$ for incident photon energies up to 20 eV with electric field polarization vector along [100] directionin details for the first time. We have used a plasma frequency of 10 eV, a Drude damping term of 0.05 eV and a Gaussian smearing of 0.5 eV for the optical property calculations of SrSi$_2$. Figure 5 displays the estimated optical properties of SrSi$_2$.

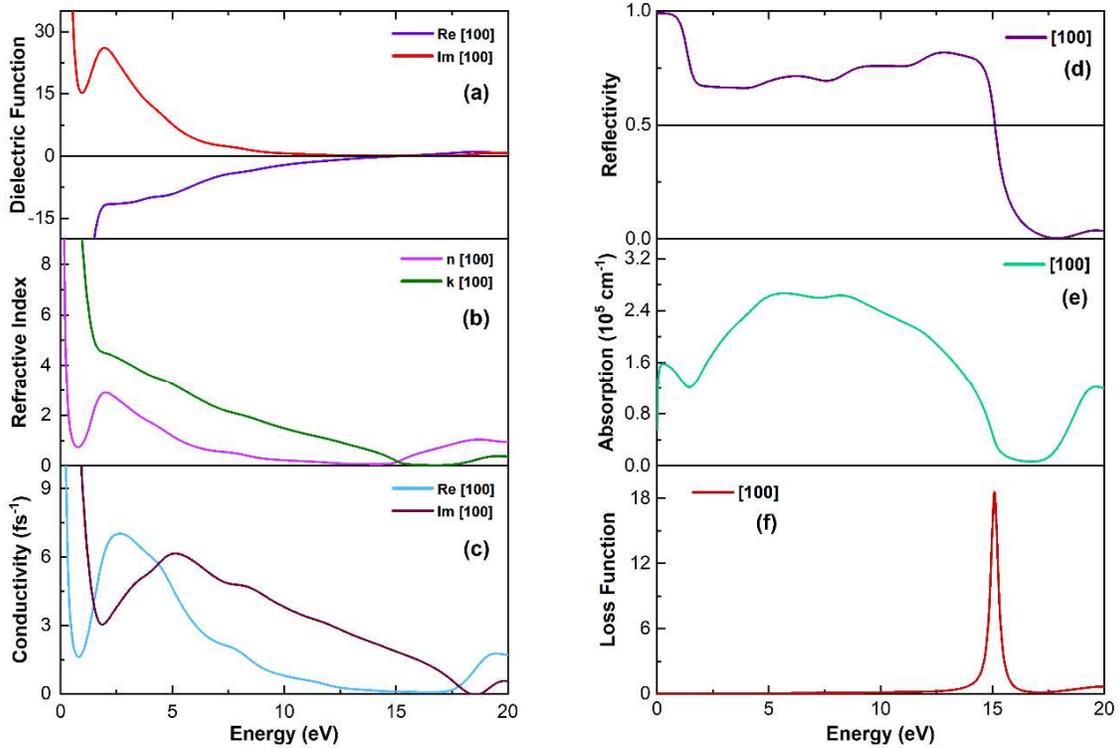

**FIG 5.** The energy dependent (a) dielectric function (real & imaginary), (b) refractive index (real & imaginary), (c) optical conductivity (real & imaginary), (d) reflectivity, (e) absorption coefficient, and (f) loss function of *SrSi$_2$*.

The real (Re) and the imaginary (Im) parts of the complex dielectric function ($\varepsilon$) of SrSi$_2$ is illustrated in figure 5(a). The polarizability of a material depends on the real part of $\varepsilon$ while the imaginary part, signifying loss, is linked with the electronic band structure and density of states (DOS). The dielectric function has contributions from both the photon-induced intra-band and inter-band transitions. The electronic band structure and peaks in the DOS plot largely control the inter-band contribution in $\varepsilon$ [89]. Thus, we anticipate that the broad peak in the imaginary part of $\varepsilon$ at around 2 eV arises as a consequence of the existence of the bonding and anti-bonding peaks in the TDOS spectrum as displayed in figure 3.The large negative value of the real part at low energies suggests the metallic character of SrSi$_2$. Since both the real and imaginary parts of $\varepsilon$ become zero at around 15 eV, we predict the energy corresponding to the plasma frequency of SrSi$_2$ is about 15 eV. Therefore, for photon energies above 15 eV SrSi$_2$ becomes fairly transparent.

Figure 5(b) depicts the real, $n(\omega)$ and the imaginary, $k(\omega)$ components of the refractive index. $n(\omega)$ is associated with the phase velocity of electromagnetic waves inside the compound. In case of SrSi$_2$, the value of $n(\omega)$ is very high in the visible region indicating its suitability to be deployed in optoelectronic display devices like LCDs, OLEDs and quantum dot (QDLED) televisions. On the other hand, $k(\omega)$, linked with the absorption coefficient, measures the attenuation of the electromagnetic wave while traversing the compound. $k(\omega)$ becomes zero at around the plasma frequency implying almost unimpeded transmission of electromagnetic energy above this particular frequency.

The metallic nature of SrSi$_2$ is evident from the optical conductivity ($\sigma$) spectra as displayed in figure 5(c). The finite value of $\sigma$ at zero photon energy indicates its metallic nature. This conclusion strongly supports the electronic band structure and DOS calculations. The optical conductivity is a measure of the conductivity of the free charge carriers due to incident photon energy. The peaks in the optical conductivity plot of SrSi$_2$ may arise because of the inter-band transitions.

In figure 5(d), the reflectivity spectrum of SrSi$_2$ is presented. It shows that the reflectivity, $R(\omega)$ of SrSi$_2$, falls sharply at the predicted plasma frequency, i.e., at 15 eV. A high value of the reflectivity (100% near 0 eV) across a broad spectrum of energy is observed in the reflectivity spectrum. This suggests that SrSi$_2$ can be considered as an efficient candidate for manufacturing optical devices where high reflectivity over a broad band is required.

The absorption coefficient, $\alpha(\omega)$, of a material is a vital quantity to determine its efficiency as a photon absorber. The absorption coefficient of SrSi$_2$ is depicted in figure 5(e). The non-zero value of $\alpha(\omega)$ at zero photon energy indicates its metallic nature which has been predicted in many other property calculations in this work. The value of $\alpha(\omega)$ remains quite high in the spectral region: 5 eV to 9 eV. Therefore, the material under study can be used as an absorber of ultraviolet light. The value of $\alpha(\omega)$ approaches zero at about 15 eV, indicating the location of plasma frequency of SrSi$_2$.

The plasma frequency is defined as the frequency at which the peak of the loss function spectrum occurs. The peak of the loss function, $L(\omega)$ of SrSi$_2$ is found at 15 eV, as illustrated in figure 5(f), consistent with the spectral features of all other optical parameters. The absorption of photon energy that leads to the plasma resonance is the reason behind the peak in the spectrum of loss function [90]. The energy loss of an electron while travelling through a material can be described using the loss function. The plasma frequency at 15 eV implies that the material under study will become transparent for photon energies above 15 eV and exhibit insulator like optical features.

## 4. Conclusions

We have studied a number of physical properties of binary silicide Weyl semimetal SrSi$_2$ in details in this work. The electronic band structure exhibits clear topological features characterized by the double Weyl nodes. The energy dispersion curves are anisotropic indicating varying career effective masses along different directions within the crystal. The Fermi level is located in the middle of a deep pseudogap separating the bonding and anti-bonding peaks in the TDOS. The Fermi surface contains both electron- and hole-like sheets. The computed band structure shows good agreement with previous studies [20,21,24]. Many of the elastic and mechanical parameters are studied for the first time. SrSi$_2$ is found to be a soft, mechanically stable, moderately machinable ductile compound. We predict significant metallic bonding in this material. The thermal parameters show close correspondence with the mechanical and bonding characters. Both Debye and melting temperatures of SrSi$_2$ are low. The compound under investigation has nonselective and high reflectivity over a broad spectral range. SrSi$_2$ is an efficient absorber of ultraviolet photons. The refractive index at low energy is high. The material has high potential to be used as a reflecting coating of solar radiation.

To summarize, we have presented a large number of novel results on various physical properties of SrSi$_2$WSM. We hope that our results will inspire the materials scientists to probe into this intriguing system in greater detail, both experimentally and theoretically.

## Acknowledgements

S. H. N. acknowledges the research grant (1151/5/52/RU/Science-07/19-20) from the Faculty of Science, University of Rajshahi, Bangladesh, which partly supported this work.

## Data availability

The data sets generated and/or analyzed in this study are available from the corresponding author on reasonable request.

## Author Contributions

S. B. performed the theoretical calculations, contributed to the analysis and draft manuscript writing. B.R.R. performed the theoretical calculations, contributed to the analysis, and contributed to manuscript writing. I.M.S. supervised the project and contributed to finalizing the manuscript. S.H.N. supervised the project, analyzed the results and finalized the manuscript. All the authors reviewed the manuscript.

## Competing Interests

The authors declare no competing interests.

# References

[1] H. B. Nielsen and M. Ninomiya, *The Adler-Bell-Jackiw Anomaly and Weyl Fermions in a Crystal*, Physics Letters B **130**, 389 (1983).

[2] A. P. Schnyder, S. Ryu, A. Furusaki, and A. W. W. Ludwig, *Classification of Topological Insulators and Superconductors in Three Spatial Dimensions*, Physical Review B **78**, 195125 (2008).

[3] A. Bansil, H. Lin, and T. Das, *Colloquium: Topological Band Theory*, Reviews of Modern Physics **88**, 021004 (2016).

[4] M. Z. Hasan, S.-Y. Xu, and G. Bian, *Topological Insulators, Topological Superconductors and Weyl Fermion Semimetals: Discoveries, Perspectives and Outlooks*, PhysicaScripta**T164**, 014001 (2015).

[5] T. Zhang, P. Cheng, X. Chen, J.-F. Jia, X. Ma, K. He, L. Wang, H. Zhang, X. Dai, Z. Fang, X. Xie, and Q.-K. Xue, *Experimental Demonstration of Topological Surface States Protected by Time-Reversal Symmetry*, Physical Review Letters **103**, 266803 (2009).

[6] P. Roushan, J. Seo, C. v. Parker, Y. S. Hor, D. Hsieh, D. Qian, A. Richardella, M. Z. Hasan, R. J. Cava, and A. Yazdani, *Topological Surface States Protected from Backscattering by Chiral Spin Texture*, Nature **460**, 1106 (2009).

[7] X.-L. Qi and S.-C. Zhang, *Topological Insulators and Superconductors*, Reviews of Modern Physics **83**, 1057 (2011).

[8] M. Z. Hasan and C. L. Kane, *Colloquium: Topological Insulators*, Reviews of Modern Physics **82**, 3045 (2010).

[9] F. D. M. Haldane, *Model for a Quantum Hall Effect without Landau Levels: Condensed-Matter Realization of the "Parity Anomaly,"* Physical Review Letters **61**, 2015 (1988).

[10] C. L. Kane and E. J. Mele, *Quantum Spin Hall Effect in Graphene*, Physical Review Letters **95**, 226801 (2005).

[11] L. Fu, C. L. Kane, and E. J. Mele, *Topological Insulators in Three Dimensions*, Physical Review Letters **98**, 106803 (2007).

[12] A. A. Burkov, *Topological Semimetals*, Nature Materials **15**, 1145 (2016).

[13] S. Jia, S.-Y. Xu, and M. Z. Hasan, *Weyl Semimetals, Fermi Arcs and Chiral Anomalies*, Nature Materials **15**, 1140 (2016).

[14] H. Weyl, *GRAVITATION AND THE ELECTRON*, Proceedings of the National Academy of Sciences **15**, 323 (1929).

[15] S.-Y. Xu, I. Belopolski, N. Alidoust, M. Neupane, G. Bian, C. Zhang, R. Sankar, G. Chang, Z. Yuan, C.-C. Lee, S.-M. Huang, H. Zheng, J. Ma, D. S. Sanchez, B. Wang, A. Bansil, F. Chou, P. P. Shibayev, H. Lin, S. Jia, and M. Z. Hasan, *Discovery of a Weyl Fermion Semimetal and Topological Fermi Arcs*, Science **349**, 613 (2015).

[16] B. Q. Lv, H. M. Weng, B. B. Fu, X. P. Wang, H. Miao, J. Ma, P. Richard, X. C. Huang, L. X. Zhao, G. F. Chen, Z. Fang, X. Dai, T. Qian, and H. Ding, *Experimental Discovery of Weyl Semimetal TaAs*, Physical Review X **5**, 031013 (2015).

[17] X. Wang, X. Pan, M. Gao, J. Yu, J. Jiang, J. Zhang, H. Zuo, M. Zhang, Z. Wei, W. Niu, Z. Xia, X. Wan, Y. Chen, F. Song, Y. Xu, B. Wang, G. Wang, and R. Zhang, *Evidence of Both Surface and Bulk Dirac Bands and Anisotropic Nonsaturating Magnetoresistance in ZrSiS*, Advanced Electronic Materials **2**, 1600228 (2016).

[18] J. Hu, S.-Y. Xu, N. Ni, and Z. Mao, *Transport of Topological Semimetals*, Annual Review of Materials Research **49**, 207 (2019).

[19] Y. Qi, P. G. Naumov, M. N. Ali, C. R. Rajamathi, W. Schnelle, O. Barkalov, M. Hanfland, S.-C. Wu, C. Shekhar, Y. Sun, V. Süß, M. Schmidt, U. Schwarz, E. Pippel, P. Werner, R. Hillebrand, T. Förster, E. Kampert, S. Parkin, R. J. Cava, C. Felser, B. Yan, and S. A. Medvedev, *Superconductivity in Weyl Semimetal Candidate $MoTe_2$*, Nature Communications **7**, 11038 (2016).

[20] S.-M. Huang, S.-Y. Xu, I. Belopolski, C.-C. Lee, G. Chang, T.-R. Chang, B. Wang, N. Alidoust, G. Bian, M. Neupane, D. Sanchez, H. Zheng, H.-T. Jeng, A. Bansil, T. Neupert, H. Lin, and M. Z. Hasan, *New Type of Weyl Semimetal with Quadratic Double Weyl Fermions*, Proceedings of the National Academy of Sciences **113**, 1180 (2016).

[21] B. Sadhukhan and T. Nag, *Electronic Structure and Unconventional Nonlinear Response in Double Weyl Semimetal $SrSi_2$*, Physical Review B **104**, 245122 (2021).

[22] G. E. Pringle, *The Structure of $SrSi_2$: A Crystal of Class O(432)*, Acta Crystallographica Section B Structural Crystallography and Crystal Chemistry **28**, 2326 (1972).

[23] J. Evers, *Transformation of Three-Dimensional Three-Connected Silicon Nets in $SrSi_2$*, Journal of Solid State Chemistry **24**, 199 (1978).

[24] B. Singh, G. Chang, T.-R. Chang, S.-M. Huang, C. Su, M.-C. Lin, H. Lin, and A. Bansil, *Tunable Double-Weyl Fermion Semimetal State in the $SrSi_2$ Materials Class*, Scientific Reports **8**, 10540 (2018).

[25] Z. Huang, Z. Chen, B. Zheng, and H. Xu, *Three-Terminal Weyl Complex with Double Surface Arcs in a Cubic Lattice*, Npj Computational Materials **6**, 87 (2020).

[26] Y.-K. Kuo, B. Ramachandran, and C.-S. Lue, *Optimization of Thermoelectric Performance of $SrSi_2$-Based Alloys via the Modification in Band Structure and Phonon-Point-Defect Scattering*, Frontiers in Chemistry **2**, (2014).

[27] K. Hashimoto, K. Kurosaki, Y. Imamura, H. Muta, and S. Yamanaka, *Thermoelectric Properties of $BaSi_2$, $SrSi_2$, and LaSi*, Journal of Applied Physics **102**, 063703 (2007).

[28] D. Shiojiri, T. Iida, T. Kadono, M. Yamaguchi, T. Kodama, S. Yamaguchi, S. Takahashi, Y. Kayama, K. Hiratsuka, M. Imai, N. Hirayama, and Y. Imai, *Re-Evaluation of the Electronic Structure and Thermoelectric Properties of Narrow-Gap Semiconducting α-$SrSi_2$: A Complementary Experimental and First-Principles Hybrid-Functional Approach*, Journal of Applied Physics **129**, 115101 (2021).

[29] Z. J. Chen and D. B. Tian, *First-Principles Calculations of Electronic, Optical, and Thermodynamic Properties of SrSi$_2$*, Journal of Applied Physics **109**, 033506 (2011).

[30] S. J. Clark, M. D. Segall, C. J. Pickard, P. J. Hasnip, M. I. J. Probert, K. Refson, and M. C. Payne, *First Principles Methods Using CASTEP*, ZeitschriftFürKristallographie - Crystalline Materials **220**, 567 (2005).

[31] R. G. Parr, *Density Functional Theory*, Annual Review of Physical Chemistry **34**, 631 (1983).

[32] W. Kohn and L. J. Sham, *Self-Consistent Equations Including Exchange and Correlation Effects*, Physical Review **140**, A1133 (1965).

[33] J. P. Perdew and A. Zunger, *Self-Interaction Correction to Density-Functional Approximations for Many-Electron Systems*, Physical Review B **23**, 5048 (1981).

[34] J. P. Perdew, K. Burke, and M. Ernzerhof, *Generalized Gradient Approximation Made Simple*, Physical Review Letters **77**, 3865 (1996).

[35] D. R. Hamann, M. Schlüter, and C. Chiang, *Norm-Conserving Pseudopotentials*, Physical Review Letters **43**, 1494 (1979).

[36] D. D. Koelling and B. N. Harmon, *A Technique for Relativistic Spin-Polarised Calculations*, Journal of Physics C: Solid State Physics **10**, 3107 (1977).

[37] T. H. Fischer and J. Almlof, *General Methods for Geometry and Wave Function Optimization*, The Journal of Physical Chemistry **96**, 9768 (1992).

[38] H. J. Monkhorst and J. D. Pack, *Special Points for Brillouin-Zone Integrations*, Physical Review B **13**, 5188 (1976).

[39] M. I. Naher and S. H. Naqib, *Structural, Elastic, Electronic, Bonding, and Optical Properties of Topological CaSn$_3$ Semimetal*, Journal of Alloys and Compounds **829**, 154509 (2020).

[40] B. R. Rano, I. M. Syed, and S. H. Naqib, *Ab Initio Approach to the Elastic, Electronic, and Optical Properties of MoTe$_2$ Topological Weyl Semimetal*, Journal of Alloys and Compounds **829**, 154522 (2020).

[41] M. I. Naher and S. H. Naqib, *An Ab-Initio Study on Structural, Elastic, Electronic, Bonding, Thermal, and Optical Properties of Topological Weyl Semimetal TaX (X = P, As)*, Scientific Reports **11**, 5592 (2021).

[42] N. Sadat Khan, B. Rahman Rano, I. M. Syed, R. S. Islam, and S. H. Naqib, *First-Principles Prediction of Pressure Dependent Mechanical, Electronic, Optical, and Superconducting State Properties of NaC$_6$: A Potential High-T$_c$ Superconductor*, Results in Physics **33**, 105182 (2022).

[43] S. Saha, T. P. Sinha, and A. Mookerjee, *Electronic Structure, Chemical Bonding, and Optical Properties of Paraelectric BaTiO$_3$*, Physical Review B **62**, 8828 (2000).

[44] M. Born, *On the Stability of Crystal Lattices. I*, Mathematical Proceedings of the Cambridge Philosophical Society **36**, 160 (1940).

[45] F. Mouhat and F.-X. Coudert, *Necessary and Sufficient Elastic Stability Conditions in Various Crystal Systems*, Physical Review B **90**, 224104 (2014).

[46] M. I, Naher, M. Mahamudujjaman, A. Tasnim, R. S. Islam, and S. H. Naqib, *Ab-initio insights into the elastic, bonding, phonon, optoelectronic and thermophysical properties of $SnTaS_2$*, arXiv:2108.07411 (2021).

[47] M. A. Ali, M. M. Hossain, M. M. Uddin, A. K. M. A. Islam, D. Jana, S. H. Naqib, *DFT insights into new B-containing 212 MAX phases: $Hf_2AB_2$ (A = In, Sn)*, Journal of Alloys and Compounds **860**, 158408 (2021).

[48] J. Worgull, E. Petti, and J. Trivisonno, *Behavior of the Elastic Properties near an Intermediate Phase Transition in $Ni_2MnGa$*, Physical Review B **54**, 15695 (1996).

[49] W. A. Harrison, *Electronic Structure and the Properties of Solids: The Physics of the Chemical Bond* (Courier Corporation, 2012).

[50] L. Kleinman, *Deformation Potentials in Silicon. I. Uniaxial Strain*, Physical Review **128**, 2614 (1962).

[51] F. Sultana, M. M. Uddin, M. A. Ali, M. M. Hossain, S. H. Naqib, and A. K. M. A. Islam, *First principles study of $M_2InC$ (M = Zr, Hf and Ta) MAX phases: the effect of M atomic species*, Results in Physics **11**, 869 (2018).

[52] M. M Hossain and S. H. Naqib, *Structural, elastic, electronic, and optical properties of layered TiNX (X = F, Cl, Br, I) compounds: a density functional theory study*, Molecular Physics **118** (3), e1609706 (2020).

[53] W. Feng and S. Cui, *Mechanical and Electronic Properties of $Ti_2AlN$ and $Ti_4AlN_3$: A First-Principles Study*, Canadian Journal of Physics **92**, 1652 (2014).

[54] D. G. Pettifor, *Theoretical Predictions of Structure and Related Properties of Intermetallics*, Materials Science and Technology **8**, 345 (1992).

[55] M. Mattesini, R. Ahuja, and B. Johansson, *Cubic $Hf_3N_4$ and $Zr_3N_4$: A Class of Hard Materials*, Physical Review B **68**, 184108 (2003).

[56] M. Jamal, S. JalaliAsadabadi, I. Ahmad, and H. A. Rahnamaye Aliabad, *Elastic Constants of Cubic Crystals*, Computational Materials Science **95**, 592 (2014).

[57] P. Ravindran, L. Fast, P. A. Korzhavyi, B. Johansson, J. Wills, and O. Eriksson, *Density Functional Theory for Calculation of Elastic Properties of Orthorhombic Crystals: Application to $TiSi_2$*, Journal of Applied Physics **84**, 4891 (1998).

[58] W. Voigt, *Lehrbuch Der Kristallphysik (Textbook of Crystal Physics)*, BG Teubner, Leipzig Und Berlin (1928).

[59] A. Reuss, *Berechnung Der Fließgrenze von Mischkristallen Auf Grund Der PlastizitätsbedingungFürEinkristalle.*, ZAMM - ZeitschriftFürAngewandteMathematik Und Mechanik **9**, 49 (1929).

[60] R. Hill, *The Elastic Behaviour of a Crystalline Aggregate*, Proceedings of the Physical Society. Section A **65**, 349 (1952).

[61] M. I. Naher, M. A. Afzal, and S. H. Naqib, *A Comprehensive DFT Based Insights into the Physical Properties of Tetragonal Superconducting $Mo_5PB_2$*, Results in Physics **28**, 104612 (2021).

[62] M. I. Naher and S. H. Naqib, *A Comprehensive Study of the Thermophysical and Optoelectronic Properties of $Nb_2P_5$ via Ab-Initio Technique*, Results in Physics **28**, 104623 (2021).

[63] S. F. Pugh, *XCII. Relations between the Elastic Moduli and the Plastic Properties of Polycrystalline Pure Metals*, The London, Edinburgh, and Dublin Philosophical Magazine and Journal of Science **45**, 823 (1954).

[64] G. N. Greaves, A. L. Greer, R. S. Lakes, and T. Rouxel, *Poisson's Ratio and Modern Materials*, Nature Materials **10**, 823 (2011).

[65] J. Cao and F. Li, *Critical Poisson's Ratio between Toughness and Brittleness*, Philosophical Magazine Letters **96**, 425 (2016).

[66] O. L. Anderson and H. H. Demarest, *Elastic Constants of the Central Force Model for Cubic Structures: Polycrystalline Aggregates and Instabilities*, Journal of Geophysical Research **76**, 1349 (1971).

[67] E. Schreiber, *Elastic Constants and Their Measurement*, Vol. 6 (1973).

[68] O. L. Anderson, *A Simplified Method for Calculating the Debye Temperature from Elastic Constants*, Journal of Physics and Chemistry of Solids **24**, 909 (1963).

[69] M. I. Naher and S. H. Naqib, *First-principles insights into the mechanical, optoelectronic, thermophysical, and lattice dynamical properties of binary topological semimetal $BaGa_2$*, arXiv:2201.00172 (2022).

[70] F. Semari, R. Boulechfar, F. Dahmane, A. Abdiche, R. Ahmed, S. H. Naqib, A. Bouhemadou, R. Khenata, X. T. Wang, *Phase stability, mechanical, electronic and thermodynamic properties of the $Ga_3Sc$ compound: an ab-initio study*, Inorganic Chemistry Communications **122**, 108304 (2020).

[71] R. Boulechfar, A. Trad Khodja, Y. Khenioui, H. Meradji, S. Drablia, Z. Chouahda, S. Ghemid, S. H. Naqib, R. Khenata, X. T, Wang, *First-principle study of the structural, mechanical, electronic and thermodynamic properties of intermetallic compounds: $Pd_3M$ (M = Sc, Y)*, International Journal of Modern Physics B **33** (27), 1950321 (2019).

[72] M. I. Naher, F. Parvin, A. K. M. A. Islam, S. H. Naqib, *Physical properties of niobium-based intermetallics ($Nb_3B$; B = Os, Pt, Au): a DFT-based ab-initio study*, The European Physical Journal B **91** (11), 1 (2018).

[73] M. F. Ashby, P. J. Ferreira, and D. L. Schodek, *Material Classes, Structure, and Properties*, in *Nanomaterials, Nanotechnologies and Design* (Elsevier, 2009), pp. 87–146.

[74] G. A. Slack, *The Thermal Conductivity of Nonmetallic Crystals*, in *Solid State Physics*, edited by H. Ehrenreich, F. Seitz, and D. Turnbull, Vol. 34 (Academic Press, 1979), pp. 1–71.

[75] M. E. Fine, L. D. Brown, and H. L. Marcus, *Elastic Constants versus Melting Temperature in Metals*, ScriptaMetallurgica **18**, 951 (1984).

[76] M. A. Ali, M. M. Hossain, A. K. M. A. Islam, S. H. Naqib, *Ternary boride $Hf_3PB_4$: Insights into the physical properties of the hardest possible boride MAX phase*, Journal of Alloys and Compounds 857, 158264 (2021).

[77] O. Bourgeois, D. Tainoff, A. Tavakoli, Y. Liu, C. Blanc, M. Boukhari, A. Barski, and E. Hadji, *Reduction of Phonon Mean Free Path: From Low-Temperature Physics to Room Temperature Applications in Thermoelectricity*, Comptes Rendus Physique **17**, 1154 (2016).

[78] D. R. Clarke, *Materials Selection Guidelines for Low Thermal Conductivity Thermal Barrier Coatings*, Surface and Coatings Technology **163–164**, 67 (2003).

[79] J. Häglund, A. Fernández Guillermet, G. Grimvall, and M. Körling, *Theory of Bonding in Transition-Metal Carbides and Nitrides*, Physical Review B **48**, 11685 (1993).

[80] A. Pasturel, C. Colinet, and P. Hicter, *Strong Chemical Interactions in Disordered Alloys*, Physica B+C **132**, 177 (1985).

[81] J.-H. Xu, T. Oguchi, and A. J. Freeman, *Crystal Structure, Phase Stability, and Magnetism in $Ni_3V$*, Physical Review B **35**, 6940 (1987).

[82] J.-H. Xu, T. Oguchi, and A. J. Freeman, *Solid-Solution Strengthening: Substitution of V in $Ni_3Al$ and Structural Stability of $Ni_3(Al,V)$*, Physical Review B **36**, 4186 (1987).

[83] T. Hong, T. J. Watson-Yang, X.-Q. Guo, A. J. Freeman, T. Oguchi, and J. Xu, *Crystal Structure, Phase Stability, and Electronic Structure of Ti-Al Intermetallics: $Ti_3Al$*, Physical Review B **43**, 1940 (1991).

[84] D. H. Douglass, editor, *Superconductivity in d- and f-Band Metals* (Springer US, Boston, MA, 1976).

[85] J. P. Carbotte, *Properties of Boson-Exchange Superconductors*, Reviews of Modern Physics **62**, 1027 (1990).

[86] V. L. Moruzzi, P. Oelhafen, A. R. Williams, R. Lapka, H.-J. Güntherodt, and J. Kübler, *Theoretical and Experimental Electronic Structure of Zr-Based Transition-Metal Glasses Containing Fe, Co, Ni, Cu, Rh, and Pd*, Physical Review B **27**, 2049 (1983).

[87] N. E. Christensen and D. L. Novikov, *Calculated Superconductive Properties of Li and Na under Pressure*, Physical Review B **73**, 224508 (2006).

[88] S. B. Dugdale, *Life on the Edge: A Beginner's Guide to the Fermi Surface*, PhysicaScripta **91**, 053009 (2016).

[89] M. A. Ghebouli, B. Ghebouli, T. Chihi, M. Fatmi, R. Khenata, H. R. Jappor, and S. H. Naqib, *Electronic Band Structure, Elastic, Optical and Thermodynamic Characteristic of Cubic YF3: An Ab Initio Study*, Optik **239**, 166680 (2021).

[90] X.-C. Ma, Y. Dai, L. Yu, and B.-B. Huang, *Energy Transfer in Plasmonic Photocatalytic Composites*, Light: Science & Applications **5**, e16017 (2016).28